# Sampled-Data and Harmonic Balance Analyses of Average Current-Mode Controlled Buck Converter

Chung-Chieh Fang *

## Abstract

Dynamics and stability of average current-mode control of buck converters are analyzed by sampled-data and harmonic balance analyses. An exact sampled-data model is derived. A new continuous-time model "lifted" from the sampled-data model is also derived, and has frequency response matched with experimental data reported previously. Orbital stability is studied and it is found unrelated to the ripple size of the current-loop compensator output. An unstable window of the current-loop compensator pole is found by simulations, and it can be accurately predicted by sampled-data and harmonic balance analyses. A new S plot accurately predicting the subharmonic oscillation is proposed. The S plot assists pole assignment and shows the required ramp slope to avoid instability.

*C.-C Fang is with Advanced Analog Technology, 2F, No. 17, Industry E. 2nd Rd., Hsinchu 300, Taiwan, Tel: +886-3-5633125 ext 3612, Email: fangcc3@yahoo.com

# Contents

# 1 Introduction

Subharmonic oscillation is a fast-scale instability commonly seen in DC-DC converters. The instability is generally associated with period-doubling bifurcation (PDB) [1]. Sampled-data models [2–7] can accurately predict PDB, which occurs when a sampled-data pole is at -1 in the complex plane. A sampled-data pole $p$, not *negative real*, can be mapped to a corresponding pole $\ln(p)/T$ in the average model. However, there is no *single* corresponding pole which can be mapped from a *negative real* sampled-data pole [8]. That explains why some average models cannot predict PDB.

In the past research [9–17], frequency responses of the average models show good matches with experimental data. However, these experiments are made under *specific* conditions when the sampled-data poles are not negative real. The deficiency or inaccuracy of the average model will be pronounced if there exists a negative real sampled-data pole. This deficiency can be eliminated by mapping ("lifting") the negative real pole to a *pair* of complex poles in the right half plane (RHP) and a zero [8], which *increases* the system dimension. With an additional pole, the lifted model has additional degree of freedom to have matched frequency response. This approach is adopted in this paper. An *exact* sampled-data model is first derived, then lifted to a continuous-time model. The lifted model is more accurate than the average model, and can predict PDB.

For a converter under *peak* current-mode control (PCMC), the current loop is one-dimensional. By considering the sampling effect and increasing the current loop dimension to *two*, improved average models [18, 19] can accurately predict PDB. In essence, the accuracy is improved by "lifting" the sampled-data model to increase the system dimension. In contrast, any improvement in the average model without increasing the system dimension is deficient in predicting PDB. The oscillation predicted by [20] is actually a Neimark bifurcation (slow-scale instability), not a PDB, because the predicted oscillation frequency is not half the switching frequency. A *single continuous-time* pole cannot cause system *oscillation*, while a single negative real *sampled-data* pole can. A *pair* of RHP continuous-time poles are required to cause oscillation. Without increased system dimension as in [18, 19], the average model generally cannot predict PDB.

Previous works on average current-mode control (ACMC) are generally based on average models [9–17], which can be divided into two groups. The first group [9–15] does not consider the sampling effect and is equivalent to *state-space* average models. The second group [16, 17] considers the sampling effect. Adding the sampling effect is effective for PCMC, but directly applying it to ACMC has been questioned [10]. Two reasons can be given. First, ACMC resembles voltage-mode control rather than PCMC [10], and considering the sampling effect may be unnecessary. Second, The current loop in ACMC has *more than one* pole, not just one as in PCMC. Some poles may migrate and regroup with each other [12, 14, 15]. One cannot predict which pole is real negative to be lifted. It is not accurate as in [16] to lift just one particular pole.

Harmonic balance modeling [21] is another approach to analyze switching converters and can predict PDB. The contributions of the paper are as follows. First, PDB does occur in ACMC. It even occurs in the circuit example in [16], and such a PDB instability was reported around a decade ago in [22]. The occurrence of PDB can be predicted by the sampled-data and harmonic balance models. Second, it is shown that the average model has frequency responses matched with experimental data only in those conditions when the converter does not have negative real sampled-data poles. The average model without increased system dimension fails to predict PDB. Third, a lifted model is proposed. It has frequency responses matched with experimental data, and it accurately predicts PDB even if there is a negative real sampled-data pole. Fourth, a new stability criterion based on the harmonic balance model is derived. Fifth, a new S plot is proposed. It accurately predicts PDB and greatly assists converter designers to adjust some particular parameters to avoid PDB. It also assists pole assignment.

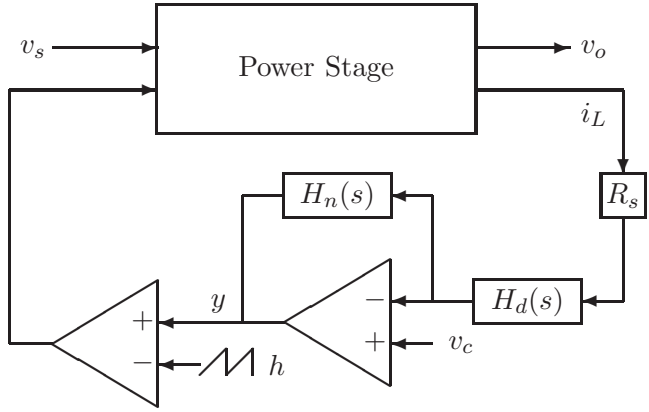

Figure 1: System diagram of an ACMC DC-DC converter.

## 2 Operation and Instability of ACMC

An ACMC DC-DC converter is shown in Fig. 1. The source voltage is $v_s$, and the output voltage is $v_o$. Let the switching frequency be $f_s = 1/T$ and let $\omega_s = 2\pi f_s$. The inductor current $i_L$ is sensed by a resistor $R_s$ and compared with a control reference $v_c$. The difference is amplified by a compensator,

$$H_c(s) = \frac{H_n(s)}{H_d(s)} = \frac{K_c(1+\frac{s}{\omega_z})}{(s+\delta)(1+\frac{s}{\omega_p})} \quad (1)$$

where $K_c$ is the compensator gain at low frequency. The mid-frequency gain is $K_c/\omega_z$. The poles are $\delta \approx 0$ and $\omega_p$. The compensator output $y(t)$ is compared with a $T$-periodic ramp signal $h(t)$, which has $h(0) = 0$ and $h(T^-) = V_h$ (ramp amplitude). The nominal solution of a DC-DC converter is a $T$-periodic orbit, not an equilibrium point as depicted in the average model. Denote the orbit by $x^0(t)$, and let $y^0(t) = Cx^0(t) + Du$. A periodic orbit (like $x^0(t)$ or $y^0(t)$) is orbitally stable if a state trajectory stays in the orbit after being perturbed. If the state trajectory eventually leaves the orbit, the orbit is unstable. Previous work [10] reported that the ripple size or slope of $y$ affects stability. However, the following simulations show otherwise. Furthermore, an unstable window of $\omega_p$ is found.

**Example 1.** (*Unstable window of $\omega_p$*.) Consider an ACMC buck converter [16, p. 114]. The system parameters are $v_s = 14$ V, $v_o = 5$ V, $v_c = 0.5$, $f_s = 50$ kHz, $L = 37.5$ $\mu$H, $C = 380$ $\mu$F with equivalent series resistance (ESR) $R_c = 0.02$ $\Omega$, $R = 1$ $\Omega$, $R_s = 0.1$ $\Omega$, $V_h$=1, $\dot{h}(d) = 50000$, $K_c = 75506$, $\omega_p = 0.492\omega_s$, and $\omega_z = 5652.9$ rad/s. All of the parameters are the same as in [16]. The converter is actually unstable [22]. The sampled-data poles are -1.123, -0.045, 0.882, and 0.9537. The pole -1.123 indicates that PDB occurs and the orbit $x^0(t)$ is unstable. When $x^0(t)$ is perturbed, the state trajectory will leave the periodic orbit and become erratic as shown in Fig. 2 and does not converge to a $T$-periodic orbit. Time simulation shows that the converter is unstable for $0.13 < \omega_p/\omega_s < 0.56$. This unstable window of $\omega_p$ can be predicted by the sampled-data and harmonic balance analyses shown later. □

**Example 2.** (*Stability is unrelated to ripple size.*) In Example 1, change $L$ to 46.1 $\mu$H, which does not change the dynamics substantially. The change of $L$ is used to illustrate a slightly smaller unstable window of $\omega_p$. Compensator pole $\omega_p$ is varied from $0.1\omega_s$ to $0.8\omega_s$. The pole $\omega_p$ adjusts how much high-order harmonics of the compensator output $y$ is attenuated. The ripple size and slope of $y$ increase as $\omega_p$ increases. An unstable *window* of $\omega_p$ was found and reported in [22]. In the window $0.18\omega_s \leq \omega_p \leq 0.49\omega_s$, $x^0(t)$ and $y^0(t)$ are unstable. Outside the window, $x^0(t)$ and $y^0(t)$ are stable. Therefore, the stability is unrelated to the ripple size: for $\omega_p < 0.18\omega_s$, the converter is stable with a small ripple size of $x^0(t)$ or $y^0(t)$; for $0.18\omega_s \leq \omega_p \leq 0.49\omega_s$, the converter is unstable with a little larger ripple size; for $\omega_p > 0.49\omega_s$, the converter is stable again with a much larger ripple size.

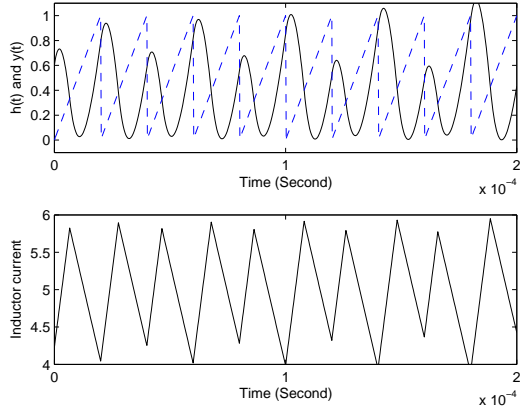

Figure 2: Unstable state trajectory after the $T$-periodic orbit $x^0(t)$ is perturbed.

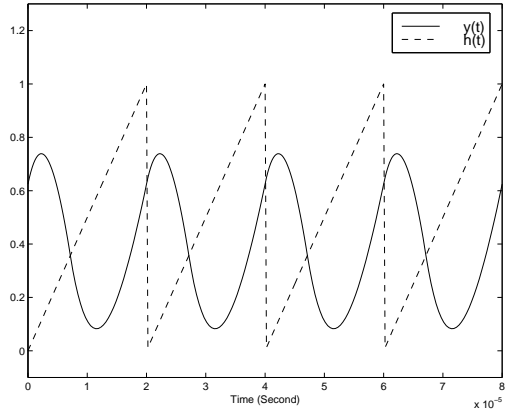

Figure 3: Unstable $T$-periodic $y^0(t)$, $\omega_p = 0.49\omega_s$.

Let $\omega_p = 0.49\omega_s$, for example. Subharmonic instability occurs, and the unstable $T$-periodic orbit (Fig. 3) and the stable $2T$-periodic orbit (Fig. 4) *coexists*. When the unstable $T$-periodic orbit is perturbed, it will lead to the stable $2T$-periodic orbit. The two orbits have the same average duty cycle 0.357 and share the same average orbits. Based on the average model, they should have the same stability. However, their stabilities are different. For $\omega_p > 0.49\omega_s$, $x^0(t)$ and $y^0(t)$ are stable. Let $\omega_p = 0.8\omega_s$, for example, $y^0(t)$ is *stable* with a *large* ripple size and slope (Fig. 5). Comparing Figs. 3 and 5, one can see that the stability is unrelated to the ripple size. It is interesting to note that as $\omega_p$ varies from $0.1\omega_s$ to $0.8\omega_s$, the average duty cycle remains the same, but the stability can be quite different. □

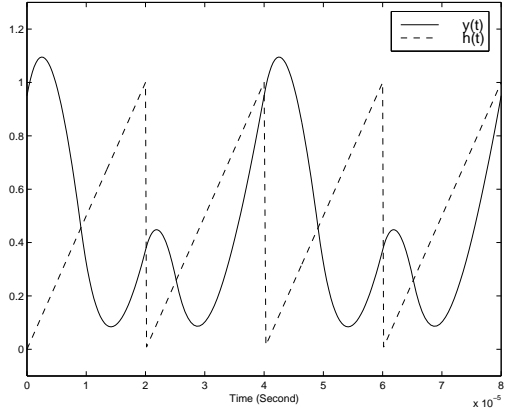

Figure 4: Stable $2T$-periodic orbit, $\omega_p = 0.49\omega_s$.

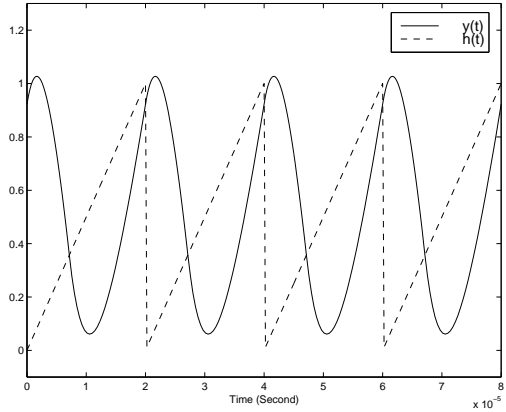

Figure 5: *Stable T-periodic* $y^0(t)$ with *large* ripple, $\omega_p = 0.8\omega_s$.

## 3 Exact General Sampled-Data Analysis

### 3.1 Exact Switching Model

The switching model [6] in Fig. 6 describes *exactly* the operation of a DC-DC converter. In the model, $A_1, A_2 \in \mathbf{R}^{N \times N}$, $B_1, B_2 \in \mathbf{R}^{N \times 2}$, $C, E_1, E_2 \in \mathbf{R}^{1 \times N}$, and $D \in \mathbf{R}^{1 \times 2}$ are constant matrices, where $N$ is the system dimension. For example, $N = 4$ for a buck converter with a two-pole compensator. Within a clock period, the dynamics is switched between $S_1$ and $S_2$. Switching occurs when the ramp signal $h(t)$ intersects with the compensator output $y := Cx + Du \in \mathbf{R}$. The system is in $S_1$ when $y(t) \geq h(t)$, and switches to $S_2$ at instants when $y(t) < h(t)$.

In the $n$-th cycle, let $x_n = x(nT)$. For high switching frequency, $u = (v_s, v_c)' \in \mathbf{R}^{2 \times 1}$ can be considered constant *within* the cycle and denoted as $u_n = (v_{sn}, v_{rn})'$. The short notation $v_{sn}$, instead of $v_{s,n}$, is used for brevity. This applies to other variables.

Let the steady-state duty cycle be $D$ and let $d = DT$. Confusion of notations for capacitance $C$ and duty cycle $D$ with the matrices $C$ and $D$ can be avoided from the context. Let $\dot{x}^0(d^-) = A_1 x^0(d) + B_1 u$ and $\dot{x}^0(d^+) = A_2 x^0(d) + B_2 u$ denote the time derivative of $x^0(t)$ at $t = d^-$ and $d^+$, respectively.

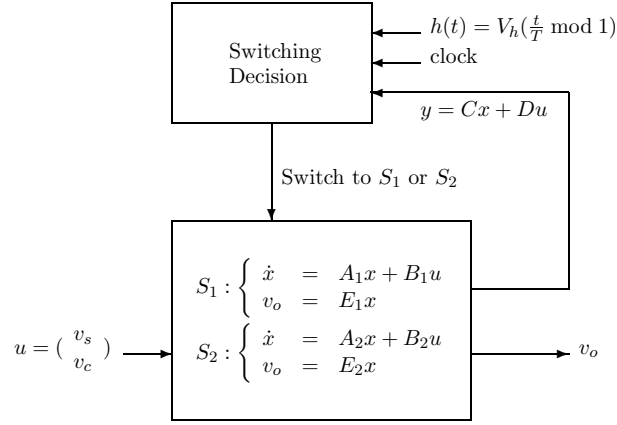

Figure 6: Exact switching model for DC-DC converter.

The periodic orbit $x^0(t)$ in the system of Fig. 6 corresponds to a fixed point $x^0(0)$ in the sampled-data dynamics. Using a hat ˆ to denote small perturbations (e.g., $\hat{x}_n = x_n - x^0(0)$). From [6], the linearized sampled-data dynamics is

$$\hat{x}_{n+1} = \Phi \hat{x}_n + \Gamma \hat{u}_n = \Phi \hat{x}_n + \Gamma_1 \hat{v}_{sn} + \Gamma_2 \hat{v}_{cn} \qquad (2)$$

where $\Phi \in \mathbf{R}^{N \times N}$ and $\Gamma = [\Gamma_1, \Gamma_2] \in \mathbf{R}^{N \times 2}$ are

$$\Phi = e^{A_2(T-d)}(I - \frac{(\dot{x}^0(d^-) - \dot{x}^0(d^+))C}{C\dot{x}^0(d^-) - \dot{h}(d)})e^{A_1 d} \qquad (3)$$

$$\Gamma = e^{A_2(T-d)}(\int_0^d e^{A_1\sigma} d\sigma B_1 - \frac{\dot{x}^0(d^-) - \dot{x}^0(d^+)}{C\dot{x}^0(d^-) - \dot{h}(d)} \cdot$$
$$(C\int_0^d e^{A_1\sigma} d\sigma B_1 + D)) + \int_0^{T-d} e^{A_2\sigma} d\sigma B_2 \qquad (4)$$

The periodic solution $x^0(t)$ is asymptotically *orbitally* stable if all of the eigenvalues of $\Phi$ are inside the unit circle of the complex plane. PDB occurs when an eigenvalue is at -1. Saddle-node bifurcation (SNB) [1] occurs when an eigenvalue is at 1.

Let $E := (E_1 + E_2)/2$, because the output voltage may be discontinuous. From (2), the control-to-output transfer function is

$$T_{oc}(z) = \frac{\hat{v}_o(z)}{\hat{v}_c(z)} = E(zI - \Phi)^{-1}\Gamma_2 \qquad (5)$$

Given a transfer function in the z domain, $T(z)$, its frequency response is $T(e^{j\omega T})$, valid up to half the switching frequency.

### 3.2 Lifted Model with Improved Frequency Response

In the sampled-data model, the switching frequency is high enough that $v_c$ is assumed constant within the cycle, which is the same as assuming a sampler with a zero-order-hold being placed after $v_c$. The sampled-data dynamics (2) can be converted back ("lifted") to a continuous-time dynamics $\dot{x} = Ax + Bu$, with the following relationship [4, 5, 8],

$$\begin{bmatrix} \Phi & \Gamma \\ 0 & 1 \end{bmatrix} = \exp(\begin{bmatrix} A & B \\ 0 & 0 \end{bmatrix} T)$$

This conversion can be done by the d2c command in Matlab. A negative real pole, if any, in the sampled-data model is mapped to a pair of complex RHP poles and a zero, and the lifted model has increased system dimension [8]. The "lifted" control-to-output transfer function is

$$E(sI - A)^{-1}B. \tag{6}$$

The "lifted" model has improved high frequency phase response because the effect of the assumption that $v_c$ being constant within the cycle becomes significant at high frequency. It will be shown (in Examples 6 and 7) that the "lifted" control-to-output transfer function has phase response matched with experimental data. The lifted model also predicts PDB.

### 3.3 "S Plot" Accurately Predicts Pole Location and PDB

Based on [4, p. 46], suppose $\lambda$ is not an eigenvalue of $e^{A_2(T-d)}e^{A_1 d}$, then $\lambda$ is an eigenvalue of $\Phi$ if and only if the following boundary condition holds:

$$\dot{y}^0(d^-) + C(\lambda e^{-A_2(T-d)}e^{-A_1 d} - I)^{-1}(\dot{x}^0(d^-) - \dot{x}^0(d^+)) = \dot{h}(d) \tag{7}$$

The proof is as follows. Suppose $\lambda$ is not an eigenvalue of $e^{A_2(T-d)}e^{A_1 d}$, then

$$\begin{aligned}
\det[\lambda I - \Phi] &= \det[\lambda I - e^{A_2(T-d)}e^{A_1 d}]\det[I + (\lambda I - e^{A_2(T-d)}e^{A_1 d})^{-1}e^{A_2(T-d)}\frac{\dot{x}^0(d^-) - \dot{x}^0(d^+)}{C\dot{x}^0(d^-) - \dot{h}(d)}Ce^{A_1 d}] \\
&= \det[\lambda I - e^{A_2(T-d)}e^{A_1 d}][1 + Ce^{A_1 d}(\lambda I - e^{A_2(T-d)}e^{A_1 d})^{-1}e^{A_2(T-d)}\frac{\dot{x}^0(d^-) - \dot{x}^0(d^+)}{\dot{y}^0(d^-) - \dot{h}(d)}]
\end{aligned}$$

$\det[\lambda I - \Phi] = 0$ requires that the last term (inside the second square brackets) of the last equation equals to zero, which leads to (7).

Since the condition (7) is in terms of signal *slopes* ($\dot{y}^0(d^-)$, $\dot{x}^0(d^-)$, $\dot{x}^0(d^+)$, and $\dot{h}(d)$), the left side of (7) is called an "S plot", $S(\lambda, D, \omega_p)$, as a function of $\lambda$, $D = d/T$, and $\omega_p$, *for example*. Then, PDB occurs when $S(-1, D, \omega_p) = \dot{h}(d)$. The converter is unstable with PDB if $S(-1, D, \omega_p) > \dot{h}(d)$. SNB occurs when $S(1, D, \omega_p) = \dot{h}(d)$. Although this paper focuses on the buck converter, the condition (7) is also applicable to other converters.

Like the popular Bode plot, the S plot facilitates the converter design. For example, the intersection of the S plot (as a function of pole $\lambda$) and $\dot{h}(d)$ shows the *real* pole location. The S plot can be used for *pole assignment*. A limitation of the pole assignment is that only *real* poles can be assigned. Choosing a particular ramp slope $\dot{h}(d)$ can assign the pole to some locations. Also, if the ramp slope $\dot{h}(d)$ is large enough such that $\dot{h}(d) > S(-1, D, \omega_p)$, PDB is avoided.

# 4 Modeling of ACMC Buck Converter

## 4.1 Sampled-Data Analysis

For the buck converter, let $x = (i_L, v_C, v_{e1}, v_{e2})'$, where $i_L$ is the inductor current, $v_C$ is the capacitor voltage, and $v_{e1}$ and $v_{e2}$ are the states of the current-loop compensator. Then,

$$A_1 = A_2 = \begin{bmatrix} \frac{-\rho R_c}{L} & \frac{-\rho}{L} & 0 & 0 \\ \frac{\rho}{C} & \frac{-\rho}{RC} & 0 & 0 \\ 0 & 0 & 0 & 1 \\ -\omega_p R_s & 0 & -\delta\omega_p & -\delta - \omega_p \end{bmatrix}$$

$$B_1 = \begin{bmatrix} \frac{1}{L} & 0 \\ 0 & 0 \\ 0 & 0 \\ 0 & \omega_p \end{bmatrix}, \quad B_2 = \begin{bmatrix} 0 & 0 \\ 0 & 0 \\ 0 & 0 \\ 0 & \omega_p \end{bmatrix}$$

$$C = \begin{bmatrix} 0 & 0 & K_c & \frac{K_c}{\omega_z} \end{bmatrix}, \quad D = \begin{bmatrix} 0 & 1 \end{bmatrix}$$

$$E_1 = E_2 = \begin{bmatrix} \rho R_c & \rho & 0 & 0 \end{bmatrix}$$

where $\rho = R/(R + R_c)$. A small nonzero $\delta$ is chosen to make both $A_1$ and $A_2$ invertible. Let $B_1 := [B_{11}, B_{12}]$, $B_2 := [B_{21}, B_{22}]$ to expand the matrices into two columns. The buck converter generally has $A_1 = A_2$, $B_{21} = 0_{N \times 1}$, and $B_{12} = B_{22}$. The boundary condition (7) becomes

$$S(\lambda, D, \omega_p) := C[(I - e^{A_1 T})^{-1}(e^{A_1 DT} - I) + I + (\lambda e^{-A_1 T} - I)^{-1}]B_{11}v_s = \dot{h}(d) \tag{8}$$

If $v_o$ is fixed, $v_s = v_o/D$ is used in the S plot.

## 4.2 Average Analysis

The linearized dynamics of the *state-space average* model for the buck converter is derived:

$$\dot{\hat{x}} = (A_1 + \frac{v_s B_{11} C}{V_h})\hat{x} + (\frac{v_s B_{11}}{V_h} + B_{12})\hat{v}_c \tag{9}$$

The control-to-output transfer function is

$$T'_{oc}(s) = \frac{\hat{v}_o(s)}{\hat{v}_c(s)} = E(sI - A_1 - \frac{v_s B_{11} C}{V_h})^{-1}(\frac{v_s B_{11}}{V_h} + B_{12}) \tag{10}$$

## 4.3 Harmonic Balance Analysis and the S Plots in Frequency Forms

Let the current loop transfer function be $T(s)$. Then, one has $T(s) = H_c(s)G_{id}(s)/V_h$, where

$$G_{id}(s) = \frac{v_s((R + R_c)Cs + 1)}{(R + R_c)LCs^2 + (L + RR_cC)s + R} \tag{11}$$

is the duty-cycle-to-inductor-current transfer function. Define a transfer function $G(s) = T(s)V_h/v_s$. From [21], PDB occurs when

$$2v_s f_s \mathbf{Re}[\sum_{k=1}^{\infty}((1 - e^{j2\pi kD})G(j\omega_s k) - G(j\omega_s(k - \frac{1}{2})))] = \dot{h}(d) \tag{12}$$

One can prove that (12) and (8) (with $\lambda = -1$) are equivalent but expressed in different *forms*, and the left side of (12) is the S plot, $S(-1, D, \omega_p)$.

Generally $G(j\omega_s)$ is low-pass, and using just one term in the summation in (12),

$$S(-1, D, \omega_p) \approx 2v_s f_s \mathbf{Re}[(1 - e^{j2\pi D})G(j\omega_s) - G(\frac{j\omega_s}{2})] \quad (13)$$

Using (1) and (11), for $\omega_s \gg 1/\sqrt{LC}$ and $1/RC$,

$$S(-1, D, \omega_p) \approx \left(\frac{4\omega_p \omega_s}{4\omega_p^2 + \omega_s^2}\right) \frac{K_c R_s v_o}{\omega_z L \pi D} \quad (14)$$

For $S(-1, D, \omega_p) > \dot{h}(d)$, instability occurs with PDB. A higher value of $S(-1, D, \omega_p)$ is likely to lead to PDB. Generally, the S plot as a function of $\omega_p$ is $\Lambda$-shaped (see Fig. 8, for example), and has a maximum value of $K_c R_s v_o / \omega_z L \pi D$ at $\omega_p = \omega_s/2$. To avoid PDB, the ramp slope $\dot{h}(d)$ should be higher than this maximum value. Therefore, a *conservative* (conservativeness because it is valid for all $\omega_p$) design guideline to avoid PDB is

$$\frac{K_c R_s v_s}{\omega_z L} = \frac{K_c R_s v_o}{\omega_z L D} < \pi \dot{h}(d) \quad (15)$$

which shows how different parameters affect stability and is useful for converter design. If the condition (15) is not met, there may exist an unstable window of $\omega_p$. Although the unstable window of $\omega_p$ predicted by (15) may be a little different from the exact condition (12), the approximate condition (15) at least predicts a characteristic of an unstable window of $\omega_p$. No existing stability criterion has been reported to predict the unstable window of $\omega_p$.

It is reported in [15] that an increase of the PWM modulator gain $1/V_h$, the mid-frequency current loop gain $K_c/\omega_z$, or $v_s$ will cause a shift of resonance frequency toward $f_s/2$. It is also reported in [15] that the load resistance $R$ does not cause such a shift. These observations agree with (15).

Note that (12) is an exact condition, (13) and (14) are approximate, and (15) is conservative. Also note that (14) is $\omega_p$-dependent, and such a dependence is removed in (15). Using which one of the conditions (12)-(15) depends on how much accuracy is required. A design guideline proposed in [12] suggests that the value of $\omega_p$ is chosen between $0.33\omega_s$ and $0.5\omega_s$. However, this range is close to the unstable window of $\omega_p$.

Another *stability* guideline proposed in [10] suggests

$$\frac{K_c R_s v_s}{\omega_z L} < \min\left[\frac{2}{1-D}, \frac{1}{D}\right] \dot{h}(d) \quad (16)$$

For a small $D$, the term $2/(1-D)$ dominates and has a value between 2 and 3. For a large $D$, the term $1/D$ dominates and has a value between 1 and 3. Either way, the right side of (16) has a value of at most 3. Therefore, (16) is more conservative than (15). The stability criterion can be relaxed as in (15) to achieve better dynamics.

## 5 Illustrative Examples

### 5.1 Many Useful Usages of the S Plot

**Example 3.** [*The S plot predicts the pole location and shows the required ramp slope to avoid PDB.*] Consider again Example 1. In Fig. 7, the intersection of the S plot (as a function of $\lambda$) and $\dot{h}(d)$ shows the two negative real poles at -1.123 and -0.045. To avoid PDB, the S plot shows that the required ramp slope is 62000 (corresponding to $V_h = 1.24$), which shifts the two poles to -0.999 and -0.051. To shift the poles from being *negative real*, the S plot also shows that the required ramp slope is 150000. With this slope, the sampled-data poles are now either complex or positive, and they are $-0.224 \pm 0.029j$, 0.872, and 0.957, agreed with the prediction of the S plot.

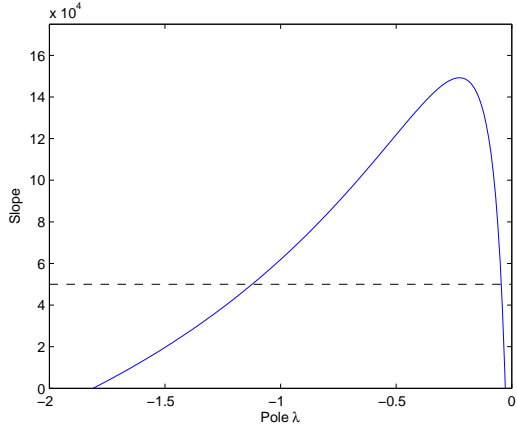

Figure 7: The intersection of $S(\lambda, 0.357, 0.492\omega_s)$ (solid line) and $\dot{h}(d)$ (dashed line) shows two poles at -1.123 and -0.045.

[*The S plot predicts the unstable window.*] In Fig. 8, the intersection of the S plot (as a function of $\omega_p$) and $\dot{h}(d)$ shows exactly the unstable window of $\omega_p$: $0.13 < \omega_p/\omega_s < 0.56$. In ACMC, the output voltage is regulated to be around $v_c$, then $D$ is varied if $v_s$ is varied. The S plot as a function of $D$ shown in Fig. 9 shows that the converter is unstable for all $D$. In both Figs. 8 and 9, a circle symbol indicates the circuit parameters (e.g., $D = 0.357$ and $\omega_p/\omega_s = 0.492$) used in Example 1. The circle symbol is above the ramp slope line at 50000, indicating instability as reported in Example 1. Both of the figures show that if the ramp slope increases to 62000 (or $V_h = 1.24$), the converter is stabilized.

With $\dot{h}(d) = 62000$, the sampled-data poles are -0.999, -0.051, 0.881, and 0.9537, indicating stability as predicted. In Fig. 9, draw a horizontal line at $\dot{h}(d) = 62000$. The line intersects with the S plot at $D = 0.35$ and 0.72. Therefore, this larger ramp slope leads to a larger stable operating range: $0.35 < D < 0.72$. The converter in Example 1 has $D = 0.357$ which is now within the stable range.

In Fig. 8, draw a horizontal line at $\dot{h}(d) = 62000$. The line intersects with the S plot at $\omega_p/\omega_s = 0.18$ and 0.49. Therefore, this larger ramp slope also leads to a smaller unstable window: $0.18 < \omega_p/\omega_s < 0.49$. The converter in Example 1 has $\omega_p = 0.492\omega_s$ which is now outside the unstable range. This window is the same as in Example 2 (with increased $L$). This can be explained by (15) because increasing $L$ or $\dot{h}(d)$ has the same effect to stabilize the converter. □

In summary, the S plot has many usages:

1. The S plot as a function of $\lambda$ (Fig. 7, for example) predicts the pole location and shows the required ramp slope to assign the *real* poles to some locations.

2. The S plot as a function of $\omega_p$ (Fig. 8, for example) predicts the stable operating range of $\omega_p$ and shows the required ramp slope to operate in a desired range of $\omega_p$ without occurrence of PDB.

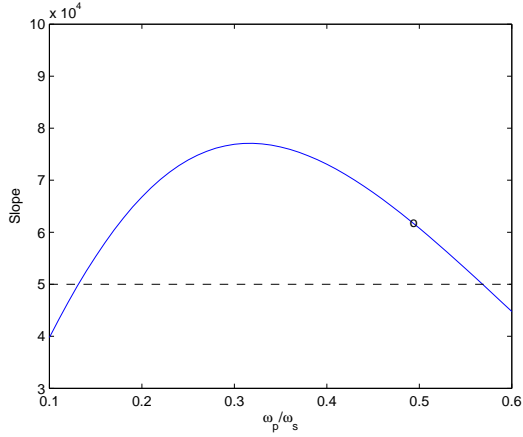

Figure 8: The intersection of $S(-1, 0.357, \omega_p)$ (solid line) and $\dot{h}(d)$ (dashed line) shows the unstable window of $\omega_p \in (0.13, 0.57)\omega_s$. The circle, for the set of converter parameters in Example 1, is above $\dot{h}(d) = 50000$, implying instability in Example 1.

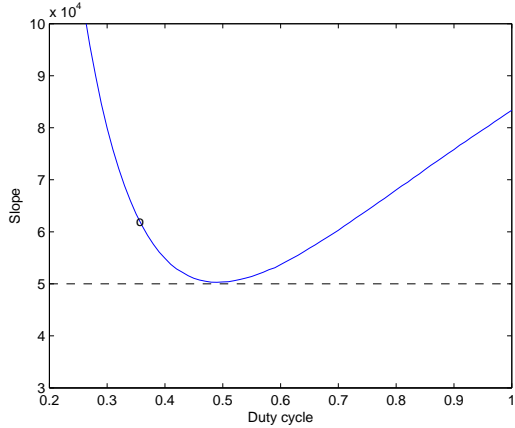

Figure 9: $S(-1, D, 0.492\omega_s)$ (solid line) is above $\dot{h}(d)$ (dashed line), showing instability for all $D$. The circle, for the set of converter parameters in Example 1, is above $\dot{h}(d) = 50000$, implying instability in Example 1.

3. The S plot as a function of $D$ (Fig. 9, for example) predicts the stable operating range of $D$ and shows the required ramp slope to operate in a desired rang of $D$ without occurrence of PDB.

If the S plot is represented as a function of another variable ($K_c$, $\omega_z$, or $L$, for example), it can predict the stable operating range of such a variable and show the required ramp slope to avoid PDB.

### 5.2 The Sample-Data Model Accurately Predicts the Unstable Window of $\omega_p$

**Example 4.** Consider again Example 1. The sampled-data model (5) is

$$\frac{0.87528(z + 0.4034)(z - 0.8987)(z - 0.0255)}{(z + 1.123)(z - 0.9537)(z - 0.882)(z + 0.04509)} \tag{17}$$

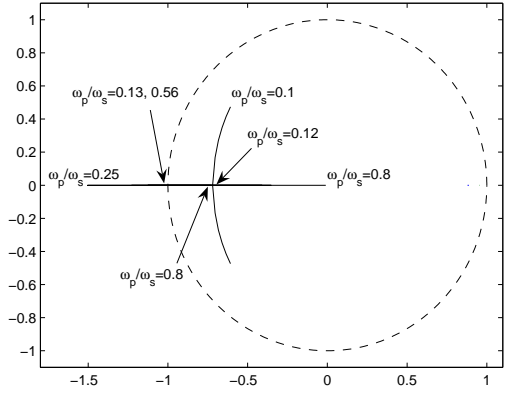

Figure 10: Sampled-data pole trajectories for $\omega_p/\omega_s \in (0.1, 0.8)$.

There are two negative real poles. One negative pole is less than -1 and is unstable. The lifted model (6) is

$$\frac{29505(s+5338)(s^2+302500s+3.812\times 10^{10})(s^2+124700s+2.981\times 10^{10})}{(s+6276)(s+2372)(s^2+309900s+4.868\times 10^{10})(s^2-11620s+2.471\times 10^{10})} \tag{18}$$

The unstable negative sampled-data poles are lifted to become a pair of unstable complex poles, accurately predicting the instability. The stable negative sampled-data poles are lifted to become a pair of stable complex poles. The lifted model now has six poles.

Next, the effect of $\omega_p$ is considered. The sampled-data pole trajectories for $\omega_p/\omega_s \in (0.1, 0.8)$ are shown in Fig. 10. Two poles are fixed around 0.882, and 0.9537 ($\approx e^{\frac{-T}{RC}}$). A pole leaves the unit circle through -1 when $\omega_p = 0.13\omega_s$, and enters the unit circle when $\omega_p = 0.56\omega_s$. This explains exactly the unstable window of $\omega_p$ discussed in Example 1. Similar techniques to increase the system dimension to predict PDB is also adopted in [23, 24]. □

**Example 5.** (*Average models do not predict PDB and the unstable window of $\omega_p$.*) In Example 1, the average model (10) is

$$\frac{7320(s+131600)(s+2216000)(s+5272)}{(s+5945)(s+2477)(s^2+149400s+7.641\times 10^{10})} \tag{19}$$

with poles in the left half plane (LHP) and fails to predict the instability.

By considering the sampling effect but *without* increasing the system dimension, another average model of [16] is

$$\frac{3.6\times 10^{14}(s+131600)}{(s+1962000)(s+154700)(s+12570)(s+1243)} \tag{20}$$

with poles still in LHP and fails to predict the instability. One explanation that these average models fail to predict PDB is that these average models do not lift the unstable sampled-data poles to increase the system dimension.

Next, the effect of $\omega_p$ is considered. The pole trajectories for $\omega_p/\omega_s \in (0.1, 0.8)$ of the average model (10) are are shown in Fig. 11. Two poles are fixed around -5945, and -2477, which can be mapped from the corresponding sampled-data poles 0.882 and 0.9537. For example, -5945≈ $\ln(0.882)/T$. All four poles are in LHP. This contradicts the unstable window in Example 1. □

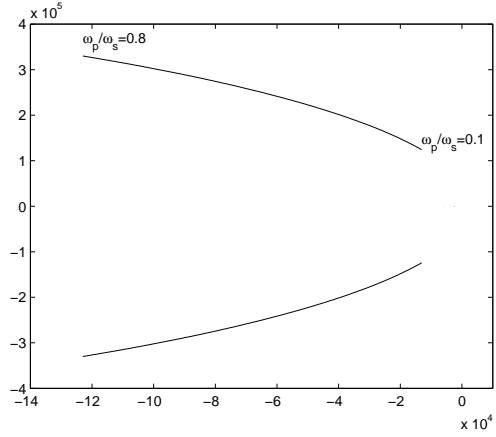

Figure 11: Average model pole trajectories for $\omega_p/\omega_s \in (0.1, 0.8)$.

## 5.3 The Lifted Model Predicts PDB and Has Matched Frequency Response

**Example 6.** Consider another ACMC buck converter from [10, p. 982]. The system parameters are $v_s = 5$ V, $v_o = 2$ V, $v_c = 0.279$, $f_s = 180$ kHz, $L = 13$ $\mu$H, $C = 750$ $\mu$F, $R_c = 5$ m$\Omega$, $R = 0.43$ $\Omega$, $R_s = 0.06$ $\Omega$, $V_h$=2.7, $\dot{h}(d) = 486000$, $K_c = 98000$, $\omega_p = \omega_s$, and $\omega_z = 6723$ rad/s.

The sampled-data model (5) is

$$\frac{0.078161(z - 0.2163)(z - 0.9654)(z + 0.08168)}{(z - 0.003783)(z - 0.5155)(z - 0.9525)(z - 0.9861)} \tag{21}$$

All four poles are positive real. Since there are no negative real poles, the average model is expected to be accurate. The lifted model (6) is

$$\frac{8100(s + 1427000)(s + 259500)(s + 6332)}{(s + 1004000)(s + 119300)(s + 8755)(s + 2528)} \tag{22}$$

There are still four poles, directly mapped from the sampled-data poles. The average model (10) is

$$\frac{704(s + 17610000)(s + 266700)(s + 6294)}{(s + 989500)(s + 133800)(s + 8523)(s + 2573)} \tag{23}$$

which has similar corresponding pole locations. The control-to-output frequency responses of the lifted and the average models are shown in Fig. 12, both matching with the experimental data shown in Fig. 6 of [10].

Here, the *average* model shows matched frequency responses. The good match in one condition does not necessarily mean that the good match occurs in all other conditions. The good match is due to the fact that here the sampled-data poles are not negative real. If one converter parameter ($v_s$, for example, as shown next) is adjusted such that the converter has a *negative real* sampled-data pole, then the average model fails to predict the converter dynamics.

The S plot (as a function of $D$) in Fig. 13 shows that PDB occurs if $D < 0.065$ (corresponding to $v_s = v_o/D = 30.84$). The approximate S plot (13) fits closely with the exact plot based on (8) or (12). The other approximate S plot (14), although not fitting with the exact plot, still accurately predicts the instability around $D = 0.065$.

Next, let $v_s = 30.84$. The unstable time waveform is similar to Figs. 2 and 4, omitted to save space. The sampled-data model (5) is

$$\frac{0.1552(z + 1.692)(z - 0.9657)(z - 0.3773)}{(z + 1.0002)(z + 0.001935)(z - 0.9623)(z - 0.9835)} \tag{24}$$

which shows PDB with two *negative real* poles. One negative pole is unstable. Lifting these two poles to totally have six poles and five zeros, with the other two stable poles being not substantially changed, the lifted model (6) is

$$\frac{137269(s + 1335000)(s + 258800)(s + 6284)(s^2 - 125700s + 2.522 \times 10^{11})}{(s^2 - 68s + 3.198 \times 10^{11})(s^2 + 2249000s + 1.584 \times 10^{12})(s + 6920)(s + 2994)} \tag{25}$$

which also has a pair of unstable poles and accurately predicts the instability. The average model (10) is

$$\frac{4343(s + 266700)(s + 6294)(s + 17610000)}{(s^2 + 1124000s + 8.62 \times 10^{11})(s + 7013)(s + 2963)} \tag{26}$$

with poles in LHP and fails to predict the instability. The control-to-output frequency responses of the lifted and the average models are shown in Fig. 14. The lifted model has a peaking at half the switching frequency and shows the instability. The average model does not show such a peaking and fails to predict the instability. Therefore, even though the average model shows matched frequency response in one condition ($v_s = 5$), it does not necessarily show accurate frequency response in another condition ($v_s = 30.84$). In contrast, in either condition, the lifted model accurately predicts the dynamics.

With $v_s = 30.84$, the S plot as a function of $\omega_p$ in Fig. 15 shows an unstable window for $0.2 \leq \omega_p/\omega_s \leq 1$. In this example, $\omega_p$ equals to $\omega_s$ and is inside this unstable window. To avoid the unstable *window*, one has two options: increase the ramp slope or operate with $\omega_p$ being outside the window.

First, consider the option of increasing the ramp slope. the S plot shows that the required slope is $\dot{h}(d) = 700000$ (corresponding to $V_h = 3.89$). Let $\dot{h}(d) = 700000$, the sampled-data poles are -0.652, -0.003, 0.962, and 0.984, indicating that PDB is avoided as expected.

Next, consider the option of operating with $\omega_p$ being outside the window. The S plot also shows that if $\omega_p/\omega_s > 1$ or $\omega_p/\omega_s < 0.2$, PDB is avoided by operating outside the unstable window. Let $\omega_p = 0.15\omega_s$, for example, the sampled-data poles are $-0.457 \pm 0.442j$, 0.962, and 0.984, indicating that PDB is avoided as expected. □

**Example 7.** Consider another ACMC buck converter from [12]. The system parameters are $v_s = 3$ V, $v_o = 2.25$ V, $v_c = 0.5625$, $f_s = 100$ kHz, $L = 20$ µH, $C = 330$ µF, $R_c = 25$ mΩ, $R = 2$ Ω, $R_s = 0.5$ Ω, $V_h = 1.8$, $\dot{h}(d) = 180000$, $K_c = 11455$, $\omega_p = 314940 \approx \omega_s/2$, and $\omega_z = 11905$ rad/s.

The sampled-data model (5) is

$$\frac{0.047467(z - 0.0746)(z - 0.9432)(z + 0.003245)}{(z^2 - 1.564z + 0.6236)(z - 0.06766)(z - 0.9889)} \tag{27}$$

All four poles are stable. Different from Example 6, a pair of complex poles exists. This example is used to show that no matter how the poles migrate and regroup, both the sampled-data and the lifted models still accurately predict the dynamics. The lifted model (6) is

$$\frac{2739(s + 273900)(s + 209300)(s + 5847)}{(s^2 + 47230s + 758500000)(s + 269300)(s + 1119)} \tag{28}$$

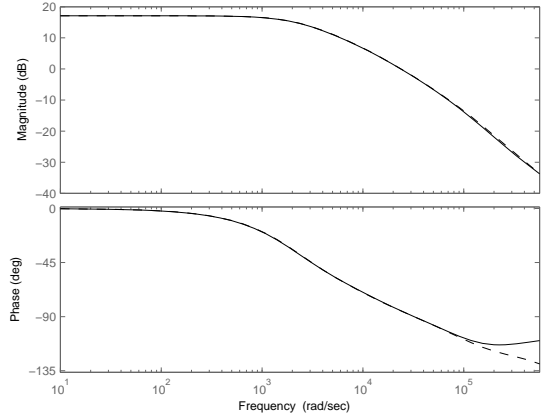

Figure 12: Control-to-output frequency responses; solid line for the lifted model and dashed line for the average model, $v_s = 5$.

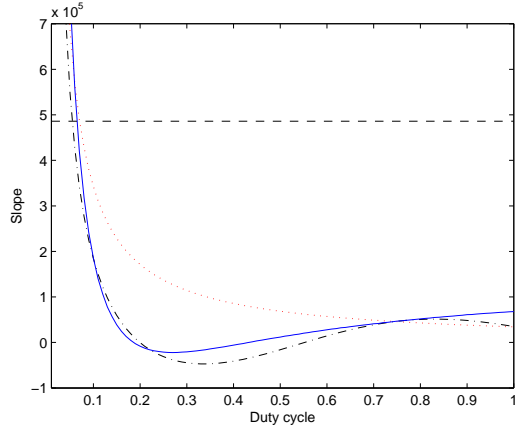

Figure 13: The intersection of $S(-1, D, \omega_s)$ (solid line) and $\dot{h}(d)$ (dashed line) shows instability for $D < 0.065$; dash-dotted line for (13) and dotted line for (14) are approximate S plots.

There are still four poles (with complex poles), mapped from the sampled-data poles. The average model (10) is

$$\frac{2058(s + 121200)(s + 612100)(s + 5894)}{(s^2 + 46410s + 745900000)(s + 270100)(s + 1116)} \tag{29}$$

which has similar corresponding pole locations. The control-to-output frequency responses of the lifted and the average models are shown in Fig. 16, both matching with the experimental data shown in Fig. 6 of [12].

Although the average model has matched frequency response, it still fails to predict PDB if $v_s$ is changed as shown next, whereas the sampled-data and the lifted models can predict PDB. The S plot in Fig. 17 shows that PDB occurs if $D < 0.09$ (corresponding to $v_s = v_o/D = 25$, also predicted by (15) based on the harmonic balance model).

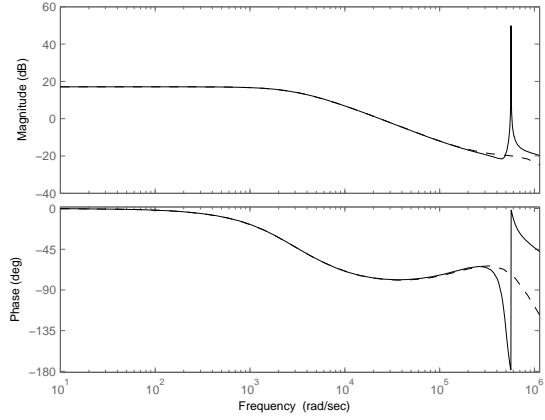

Figure 14: Control-to-output frequency responses; solid line for the lifted model and dashed line for the average model, $v_s = 30.84$.

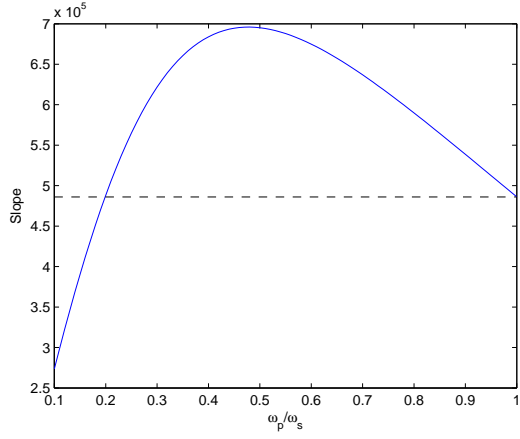

Figure 15: The intersection of $S(-1, 0.4, \omega_p)$ (solid line) and $\dot{h}(d)$ (dashed line) shows the unstable window of $\omega_p \in (0.2, 1)\omega_s$.

Next, let $v_s = 25$. The sampled-data model (5) is

$$\frac{0.28065(z - 0.9422)(z - 0.4226)(z + 0.5305)}{(z + 1.023)(z + 0.04694)(z - 0.8816)(z - 0.9856)} \tag{30}$$

which shows PDB with two *negative real* poles. Lifting these two poles, the lifted model (6) is

$$\frac{30795(s + 563100)(s + 101100)(s + 5961)(s^2 + 125600s + 1.318 \times 10^{11})}{(s^2 - 4558s + 9.87 \times 10^{10})(s^2 + 611800s + 1.923 \times 10^{11})(s + 12600)(s + 1448)} \tag{31}$$

which also has a pair of unstable poles. The average model (10) is

$$\frac{17147(s + 121200)(s + 612100)(s + 5894)}{(s^2 + 303400s + 1.02 \times 10^{11})(s + 12810)(s + 1436)} \tag{32}$$

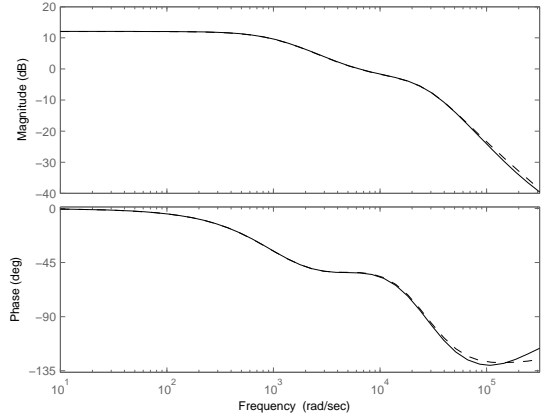

Figure 16: Control-to-output frequency responses; solid line for the lifted model and dashed line for the average model.

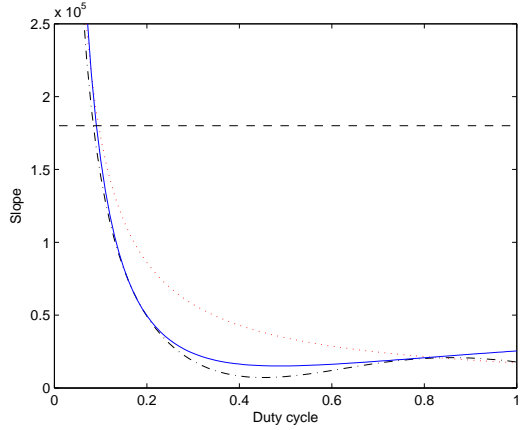

Figure 17: The intersection of $S(-1, D, \omega_s/2)$ (solid line) and $\dot{h}(d)$ (dashed line) shows instability for $D < 0.09$; dash-dotted line for (13) and dotted line for (14) are approximate S plots.

with poles in LHP and fails to predict the instability.

With $v_s = 25$, the S plot as a function of $\omega_p$ in Fig. 18 shows an unstable window for $0.36 < \omega_p/\omega_s < 0.54$. To avoid the unstable *window*, the S plot shows the required slope is $\dot{h}(d) = 185000$ (or $V_h = 1.85$).

Next, let $\dot{h}(d) = 185000$, the sampled-data poles are -0.982, -0.049, 0.881, and 0.986, indicating that PDB is avoided as expected. The S plot also shows that if $\omega_p/\omega_s > 0.54$ or $\omega_p/\omega_s < 0.36$, PDB is avoided by operating outside the unstable window. Let $\omega_p = 0.55\omega_s$, for example, the sampled-data poles are -0.991, -0.036, 0.882, and 0.986, indicating that PDB is avoided as expected. □

# 6 Conclusion

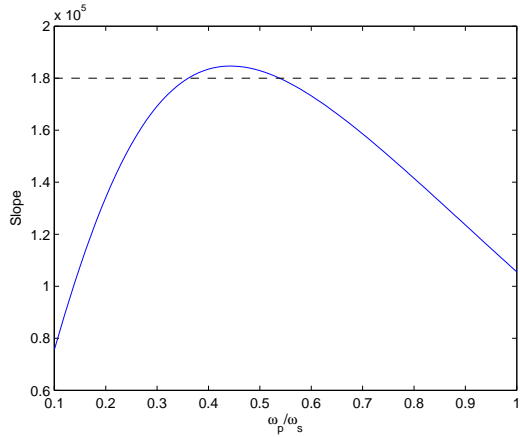

Figure 18: The intersection of $S(-1, 0.75, \omega_p)$ (solid line) and $\dot{h}(d)$ (dashed line) shows the unstable window of $\omega_p \in (0.36, 0.54)\omega_s$.

Sampled-data modeling and harmonic balance modeling are applied to analyze ACMC buck converters. In the sampled-data model, the system dynamics is derived *directly* and *exactly* from the exact switching model. The *orbital* nature of the nominal periodic solution is preserved. Orbital stability is studied and is unrelated to the ripple size of the current-loop compensator output. A new continuous-time model "lifted" from the sampled-data model is also derived, and has frequency response matched with experimental data reported previously. An unstable window of $\omega_p$ (current-loop compensator pole) which leads to PDB is found by simulations. PDB can be predicted by the sampled-data model, the lifted model and the harmonic balance model, but not the average model if the system dimension is not increased. Similar analysis can be applied to other types of converters.

Accurate PDB boundary conditions, such as (7), (8), (12)-(15) are derived. The boundary conditions, which also lead to a new "S plot", greatly assist converter design to avoid instability, just like the popular Bode plot. The S plot has many expressions: (7) is for any DC-DC converters and is exact, (8) and (12)-(15) are for buck converters, (8) is exact in terms of matrices, (12) is exact in terms of signal harmonics, and (13)-(15) are approximate.

With some converter parameters, the average model may show frequency responses matched with experimental data. The good match in one condition does not necessarily mean that the good match occurs in all other conditions. If one converter parameter is adjusted such that the converter has a *negative real* sampled-data pole, the average model without increased system dimension does not accurately predict the converter dynamics. A negative real sampled-data pole implies that the system is oscillatory. The traditional average model without increased system dimension does not have a corresponding *single* pole mapped from the negative real sampled-data pole. This deficiency is removed in the lifted model. In the lifted model, the system dimension is increased to generate additional degree of freedom to have better and matched frequency response. Therefore, if the converter does not have a negative real sampled-data pole, the average model is adequate. If the converter has a negative real sampled-data pole, the sampled-data and the lifted models will give more accurate results.

The following design procedure for ACMC is suggested. First, use (14) or (15) for an initial design. Compared with the exact condition (12), the conditions (14) and (15) are conservative, but they are not as conservative as the traditional guideline (16). Next, use the S plot (8) or (12), as a function of $D$ or $\omega_p$, to confirm that PDB is avoided. Increase the ramp slope above the S plot to have sufficient stability margin. The S plot can be also used for pole assignment. If an unstable window of $\omega_p$ exists, generally between $0.3\omega_s$ and $0.5\omega_s$, operate outside the window or adjust converter parameters according to (14), such as decrease $K_c R_s v_s$, or increase $\omega_z L$. Then, plot the frequency response of the lifted model (6) to assist the voltage-loop compensation design. Knowledge about the *sampled-data* pole locations are also useful for *digital* control of the converter.

# References

[1] C.-C. Fang and E. H. Abed, "Saddle-node bifurcation and Neimark bifurcation in PWM DC-DC converters," in *Nonlinear Phenomena in Power Electronics: Bifurcations, Chaos, Control, and Applications*, S. Banerjee and G. C. Verghese, Eds. New York: Wiley, 2001, pp. 229–240.

[2] A. Brown and R. D. Middlebrook, "Sampled-data modelling of switching regulators," in *Proc. IEEE PESC*, 1981, pp. 349–369.

[3] G. C. Verghese, M. Elbuluk, and J. G. Kassakian, "A general approach to sample-data modeling for power electronic circuits," *IEEE Trans. Power Electron.*, vol. 1, no. 2, pp. 76–89, 1986.

[4] C.-C. Fang, "Sampled-data analysis and control of DC-DC switching converters," Ph.D. dissertation, Dept. of Elect. Eng., Univ. of Maryland, College Park, 1997, available: http://www.lib.umd.edu/drum/, also published by UMI Dissertation Publishing in 1997.

[5] C.-C. Fang and E. H. Abed, "Sampled-data modeling and analysis of power stages of PWM DC-DC converters," *Int. J. of Electron.*, vol. 88, no. 3, pp. 347–369, March 2001.

[6] ——, "Sampled-data modeling and analysis of closed-loop PWM DC-DC converters," in *Proc. IEEE ISCAS*, vol. 5, 1999, pp. 110–115.

[7] C.-C. Fang, "Sampled-data modeling and analysis of one-cycle control and charge control," *IEEE Trans. Power Electron.*, vol. 16, no. 3, pp. 345–350, May 2001.

[8] I. Kollár, G. Franklin, and R. Pintelon, "On the equivalence of z-domain and s-domain models in system identification," in *IEEE Instrumentation and Measurement Technology Conference*, June 1996, pp. 14–19.

[9] L. H. Dixon, "Average current-mode control of switching power supplies," *Unitrode Power Supply Design Seminar Handbook*, 1990.

[10] J. Sun and R. M. Bass, "Modeling and practical design issues for average current control," in *Proc. IEEE APEC*, 1999, pp. 980–986.

[11] C. Sun, B. Lehman, and R. Ciprian, "Dynamic modeling and control in average current mode controlled PWM DC/DC converter," in *Proc. IEEE PESC*, 1999, pp. 1152–1157.

[12] P. Cooke, "Modeling average current mode control," in *Proc. IEEE APEC*, 2000, pp. 256–262.

[13] T. Suntio, J. Lempinen, I. Gadoura, and K. Zenger, "Dynamic effects of inductor current ripple in average current mode control," in *Proc. IEEE PESC*, 2001, pp. 1259–1264.

[14] R. Li, T. O'Brien, J. Lee, and J. Beecroft, "A unified small signal analysis of DC-DC converters with average current mode control," in *IEEE Energy Conversion Congress and Exposition*, 2009, pp. 647–654.

[15] ——, "Effects of circuit and operating parameters on the small-signal dynamics of average-current-mode-controlled DC-DC converters," in *IEEE 8th International Conference on Power Electronics and ECCE Asia*, 2011, pp. 60–67.

[16] W. Tang, F. C. Lee, and R. B. Ridley, "Small-signal modeling of average current-mode control," *IEEE Trans. Power Electron.*, vol. 8, no. 2, pp. 112–119, 1993.

[17] Y.-S. Jung, J.-Y. Lee, and M.-J. Youn, "A new small-signal modeling of averaging current mode control," in *Proc. IEEE PESC*, 1998, pp. 1118–1124.

[18] R. B. Ridley, "A new, continuous-time model for current-mode control," *IEEE Trans. Power Electron.*, vol. 6, no. 2, pp. 271–280, 1991.

[19] F. D. Tan and R. D. Middlebrook, "Unified modeling and measurement of current-programmed converters," in *Proc. IEEE PESC*, 1993, pp. 380–387.

[20] J. Sun and B. Choi, "Averaged modeling and switching instability prediction for peak current control," in *Proc. IEEE PESC*, 2005, pp. 2764–2770.

[21] C.-C. Fang and E. H. Abed, "Harmonic balance analysis and control of period doubling bifurcation in buck converters," in *Proc. IEEE ISCAS*, vol. 3, May 2001, pp. 209–212.

[22] C.-C. Fang, "Modeling and instability of average current control," in *International Power Electronics And Motion Control Conference*, 2002, paper SSIN-03.

[23] J. Li and F. C. Lee, "New modeling approach and equivalent circuit representation for current-mode control," *IEEE Trans. Power Electron.*, vol. 25, no. 5, pp. 1218–1230, 2010.

[24] F. Yu, "Modeling of $V^2$ control with composite capacitors and average current mode control," Master's thesis, Virginia Polytechnic Institute and State University, Blacksburg, VA, May 2011.